\pdfoutput=1
\documentclass[aps,prd,reprint,superscriptaddress,longbibliography,nofootinbib]{revtex4-1}
\usepackage{graphicx}
\usepackage{amsmath}
\usepackage{amssymb}
\usepackage{amsfonts}
\usepackage{dcolumn}
\usepackage{dsfont}
\usepackage{latexsym}
\usepackage{rotating}
\usepackage{color}
\usepackage{latexsym}
\usepackage{bbm}
\usepackage{subfigure}
\usepackage{float}
\usepackage{epsfig}
\usepackage{psfrag}
\usepackage{natbib, hyperref}
\usepackage{bm}
\usepackage{amsthm}
\usepackage{eucal}
\usepackage{mathrsfs}
\usepackage{url}
\usepackage{braket}
\usepackage{ulem}
\usepackage{calligra}
\usepackage[T1]{fontenc}
\usepackage[utf8]{inputenc}
\usepackage{newunicodechar}
\usepackage{marvosym}
\usepackage[absolute]{textpos}
\usepackage[cal=cm]{mathalfa}
\usepackage{soul}
\usepackage[utf8]{inputenc}

\usepackage{color}

\usepackage{hyperref}
\hypersetup{
colorlinks=true,final=true,
        linkcolor=blue,
        citecolor=blue,
        filecolor=blue,
        urlcolor=blue,
}

\begin{document}
\setlength{\abovedisplayskip}{3pt}
\setlength{\belowdisplayskip}{3pt}

\title{Anomalous magnetoentropic response of skrymion crystals}
\author{Ahmed R. Saikia}
\affiliation{Department of Physics, Indian Institute of Technology Roorkee, Roorkee 247667, India}
\author{Narayan Mohanta}
\email{narayan.mohanta@ph.iitr.ac.in}
\affiliation{Department of Physics, Indian Institute of Technology Roorkee, Roorkee 247667, India}

\begin{abstract}
We investigate theoretically magnetoentropic signatures of the crystal phase of magnetic skyrmions of various kinds, commonly appearing in two dimensions, \textit{viz.}, N\'eel, Bloch and anti skyrmions. Using Monte Carlo calculations based on spin Hamiltonians, we obtain magnetic entropy change $\Delta S_m$ in the presence of three different types of Dzyaloshinskii-Moriya interactions responsible for these skyrmions. The phase mapping of $\Delta S_m$ using skyrmion counting number $N_{sk}$ in temperature-magnetic field plane reveals fluctuation-dominated weak first-order transition in the precursor phase of the skyrmions, and a sign change in $\Delta S_m$ when the system enters into the skyrmion crystal phase---in agreement with recent experimental findings. We also find that the fractional entropy change in going from a ferromagnetic phase to the skyrmion crystal phase is much larger compared to the conventional route of paramagnetic phase to ferromagnetic phase, used for the purpose of magnetic cooling. The magnetoentropic signatures of the different types of skyrmion crystals are found to be similar. Our results indicate that the skyrmion crystals exhibit enhanced cooling efficiency and have the potential to upgrade the existing magnetic cooling methods.
\end{abstract}

\maketitle

\section{Introduction} 
\vspace{-0.3cm}
Chiral spin fluctuations in magnets with antisymmetric Dzyaloshinskii-Moriya interaction (DMI) near the onset of magnetically ordered phases give rise to anomalies in various physical observables including magnetization, specific heat, electric polarization and magnetic susceptibility~\cite{Pappas_PRL2009,Ruff_SciRep2015,Pardeep_JPCM2023}. These chiral spin fluctuations, driven primarily by the frustration introduced by the DMI and aided by intrinsic inhomogeneities, also generate strong signatures in the magnetic entropy near the temperature-driven transition to the ordered phases. In chiral magnets and spinel compounds, which support skyrmion crystal (SkX) phase at high temperatures, the magnetic entropy response was found to clearly distinguish the SkX phase from other topologically trivial phases such as conical and ferromagnetic phases~\cite{Ge_JAC2015,Han_MResBull2017,Bocarsly_PRB2018,Jamaluddin_AFM2019,Dhital_PRB2020,Zuo_PRMater2021}. The SkX phase appears also robustly in two dimensions, at the interface between two oxide compounds and in van der Waals magnets, stabilized by a magnetic anisotropy which originates from interfacial strain and favors the long-range magnetic order~\cite{Balents_PRL2014,Yokouchi_JPSJ2015,Matsuno_SciAdv2016,Rowland_PRB2016,Hoffmann2017,Nakamura_JPSJ2018,Vistoli_NPhys2019,Mohanta_PRB2019,Park_PRB2021}. Within a range of temperatures above the critical temperature for the SkX phase, the average scalar spin chirality $\chi \!=\!\langle {\bf S}_i \cdot ({\bf S}_j \times {\bf S}_k)\rangle$ can be finite even when $\langle {\bf S}_i \rangle \approx 0$. The nature of the phase transition from the paramagnetic phase to the SkX phase is, therefore, dictated primarily by the spin chirality fluctuation. 

In this paper, we investigate the behavior of the magnetic entropy as the transition to the SkX phase takes place in two dimensional magnets, and explore the classification strategy of the SkX phase using the magnetoentropic response. The central object of our investigation is the magnetic entropy change, which can be obtained using the following Maxwell's thermodynamics relation
\begin{align} 
\Big ( \frac{\partial S}{\partial H} \Big )_T=\Big ( \frac{\partial M}{\partial T} \Big )_H
\label{entropy_change}
\end{align}

The above relation provides the magnetic entropy change $\Delta S_m$ in terms of the temperature dependence of magnetization. We perform Monte Carlo calculations and obtain $\Delta S_m$ as a function of temperature and magnetic field, and obtain the magnetoentropic response of various ordered phases. Magnetic skyrmions are topologically stable spin texture which can be characterized by an invariant known as the skyrmion number, defined as
\begin{align} 
N_{sk}=\frac{1}{4\pi}\int {\bf S} \cdot \Big ( \frac{\partial {\bf S}}{\partial x} \times \frac{\partial {\bf S}}{\partial y} \Big )dx~dy.
\label{sk_num}
\end{align}
The SkX phase exhibits a large value of $N_{sk}$; hence it provides an useful tool to map out the SkX phase by comparing $N_{sk}$ with $\Delta S_m$ in the phase plane spanned by magnetic field and temperature. We perform this magnetoentropic mapping for three different types of DMI in two dimensions which give rise to three common types of skyrmions, \textit{viz.}, N\'eel, Bloch and anti skyrmions. Besides, $\Delta S_m$ can also shed light on the puzzle in understanding the role of critical spin fluctuations as the system transitions to the SkX phase when the temperature is lowered. The puzzle arises because specific heat measurements reveal a first-order transition at zero magnetic field to the helical state~\cite{Bauer_PRL2013}, while neutron probes reveal Landau soft-mode mechanism of weak crystallization to the SkX phase at finite magnetic fields~\cite{Wilhelm_PRL2011,Samatham_PSS2013,Kindervater_PRX2019}. These experimental observations suggest that phase transitions in Dzyaloshinskii-Moriya magnets occur via a precursor phase in which chiral spin textures emerge with an abundance of spin fluctuations. The magnetic entropy change $\Delta S_m$ can provide a great amount of information about the fluctuation-induced gradual first-order transition to the SkX phase~\cite{Caron_JMMM2009}. Valleys and peaks in $\Delta S_m$ can reveal phase transitions which can be detected in experiments and give information complementary to the conventional signatures in the heat capacity $C\!=\!T(\partial S/\partial T)$~\cite{Dhital_PRB2020}. We find that for all three types of skyrmions, there is a change in the slope in the temperature variation of $\Delta S_m$ as the system enters into the precursor phase from the spin-disordered paramagnetic phase. Also, $\Delta S_m$ undergoes a change in sign as the system enters the fully-ordered SkX phase. These features enable us to identify the critical temperatures for the precursor phase and the fully-ordered SkX phase. The magnetic entropy change can also provide valuable insights into entropy-limited topological protection and lifetime of skyrmions~\cite{Wild_Sciadv2017}. 

Conventional magnetic alloy compounds exhibiting a large magnetic entropy change also suffer from considerable thermal hysteresis effect which makes these materials unsuitable for applications in refrigerators due to their low performance in the cyclic operation at high cycle frequencies~\cite{MEchange}. The irreversibility of the magnetization also causes serious problems in estimating the magnetocaloric effect from the magnetization measurements~\cite{Amaral_APL2009,Balli_APL2009,Magnus_JAC2011}. We here also discuss about magnetic cooling efficiency of a cooling cycle involving a SkX phase, and compare it qualitatively with the conventional cycle involving the paramagnetic and ferromagnetic phases. Based on our estimation of a large fractional entropy change, we propose that the cyclic route between the ferromagnetic phase and the SkX phase can enable a more efficient magnetic cooling method. Compounds hosting topological spin textures such as the SkX can, therefore, be a promising alternative for the refrigeration applications with high efficiency.


The remainder of this paper is organized as follows: in section II, we discuss our theoretical model and provide details of our Monte Carlo annealing simulation. In section III, we present the numerical results of the magnetic entropy change $\Delta S_m$ as the temperature is reduced while approaching various phases {\textit viz.} spin spiral (SS), SkX and ferromagnetic (FM) phases, all stabilized at low temperatures in our considered two-dimensional spin systems. In section III, we discuss the experimental relevance of our results, enhanced magnetic cooling efficiency of the SkX phase, and summarize our results.

\section{Model and method}
We consider two-dimensional spin systems with competing Heisenberg exchange interaction and DMI, in the presence of an external magnetic field applied perpendicular to the two-dimensional plane (considered to be the $x$-$y$ plane) and an easy-plane magnetic anisotropy. The model Hamiltonian for such systems is given by
\begin{align}
\mathcal{H} = -J  \sum_{\langle ij\rangle} {\bf S}_i \cdot {\bf S}_j + \mathcal{H}_{\rm DMI} -H_z \sum_{i} S_{zi} -A\sum_{i} |S_{zi}|^2
\end{align}
where ${\bf S_i}\!\equiv \!(S_{xi},S_{yi},S_{zi})$ represents the spin vector of magnitude $S\!=\!1$ at site $i$, $J$ is the strength of nearest-neighbor Heisenberg exchange interaction, $\mathcal{H}_{\rm DMI}$ is the Hamiltonian describing the DMI which is mentioned below, $H_z$ is the strength of the perpendicular magnetic field, and $A$ is the easy-plane magnetic anisotropy which helps to stabilize the magnetic skyrmions and the long-range order in the SkX~\cite{Bogdanov_JMMM1999,Banerjee_PRX2014,Kovalev_PRB2016,Leonov_PRB2017}. We used $A\!=\!0.001J$ throughout this paper.  

\begin{figure}[t]
\begin{center}
\vspace{-0mm}
\epsfig{file=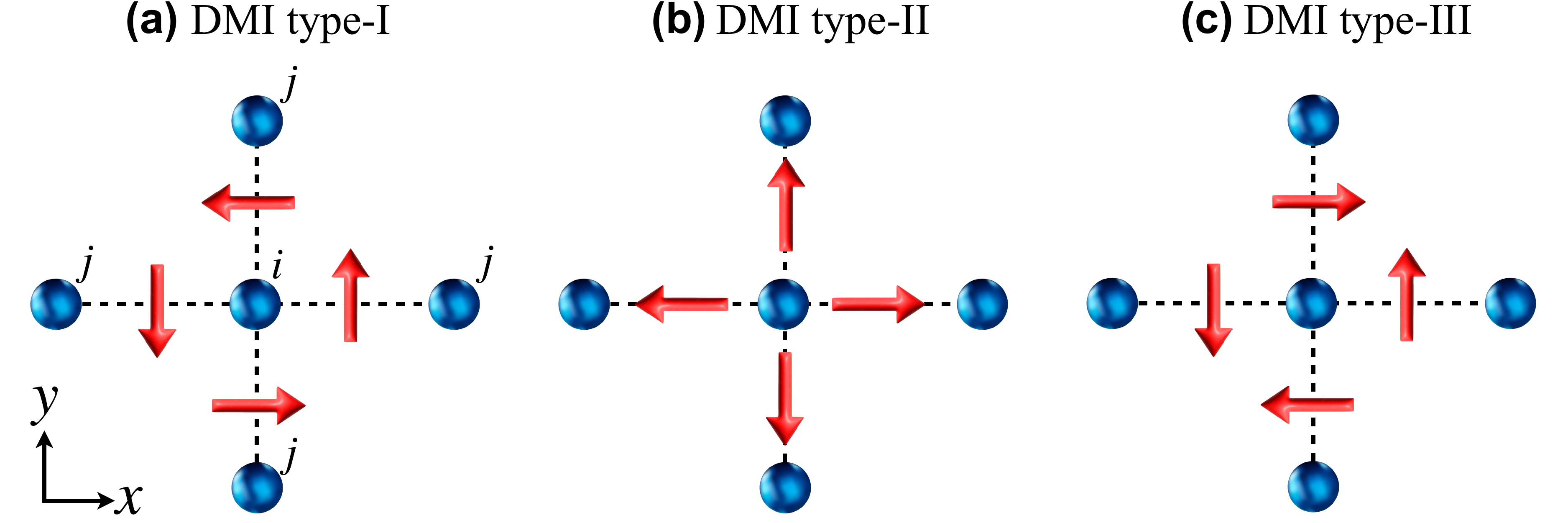,trim=0.0in 0.0in 0.0in 0.0in,clip=false, width=86mm}
\caption{Schematic representation of the three types of Dzyaloshinskii-Moriya interaction (DMI) considered in the two-dimensional geometry. Blue spheres represent nearest-neighbor lattice sites, with indices $i$ and $j$. Red arrows denote the vectors of the DMI in the three cases, which give rise to (a) N\'eel, (b) Bloch and (c) anti skyrmions, respectively.}
\label{fig1}
\vspace{-3mm}
\end{center}
\end{figure}

As discussed widely in literature, the DMI term is relevant in materials with strong spin-orbit coupling and broken inversion symmetry. Magnets with sizeable Rashba spin-orbit coupling, such as those interfacing another compound in a two-dimensional plane, are candidate systems where the DMI amplitude is typically large. Here we consider three different types of DMI, commonly appearing in two dimensions and responsible for the N\'eel, Bloch and anti skyrmions. The vectors representing these three types of DMIs are shown in Fig.~\ref{fig1}, and the explicit forms of $\mathcal{H}_{\rm DMI}$ in these three cases are expressed as
\begin{align}
&\mathcal{H}_{\rm DMI} = - D \sum_{\langle ij\rangle} (\hat{z}\times\hat{r}_{ij})\cdot({\bf S}_i\times{\bf S}_j),\\
&\mathcal{H}_{\rm DMI} = - D \sum_{\langle ij\rangle,\hat{e}=\hat{x},\hat{y}} (\pm\hat{e})\cdot({\bf S}_{i}\times{\bf S}_{j=i\pm\hat{e}}),\\
&\mathcal{H}_{\rm DMI} = -  D \sum_{\langle ij\rangle} \eta (\hat{z}\times\hat{r}_{ij})\cdot({\bf S}_i\times{\bf S}_j),
\end{align}
where, in the last expression $\eta \!=\!+1$ for $j\!=\!i\pm \hat{x}$ and $\eta \!=\!-1$ for $j\!=\!i\pm \hat{y}$. The DMI in Eq.(4), Eq.(5) and Eq.(6) stabilizes, respectively, the N\'eel, Bloch and anti skyrmions in two dimensional lattices. We consider $J\!=\!1$, and express all other energies in units of $J$.

The SkX phase can also originate without the explicit need of the DMI or external magnetic field, from spin frustration or interfacial phase frustration~\cite{Ozawa_PRL2017,Hayami_PRB2021,Mohanta_npjQM2022}. We anticipate our discussions below and the magnetoentropic signatures to be relevant in those unconventional SkXs as well.\\
\begin{figure*}[ht]
\begin{center}
\vspace{-0mm}
\epsfig{file=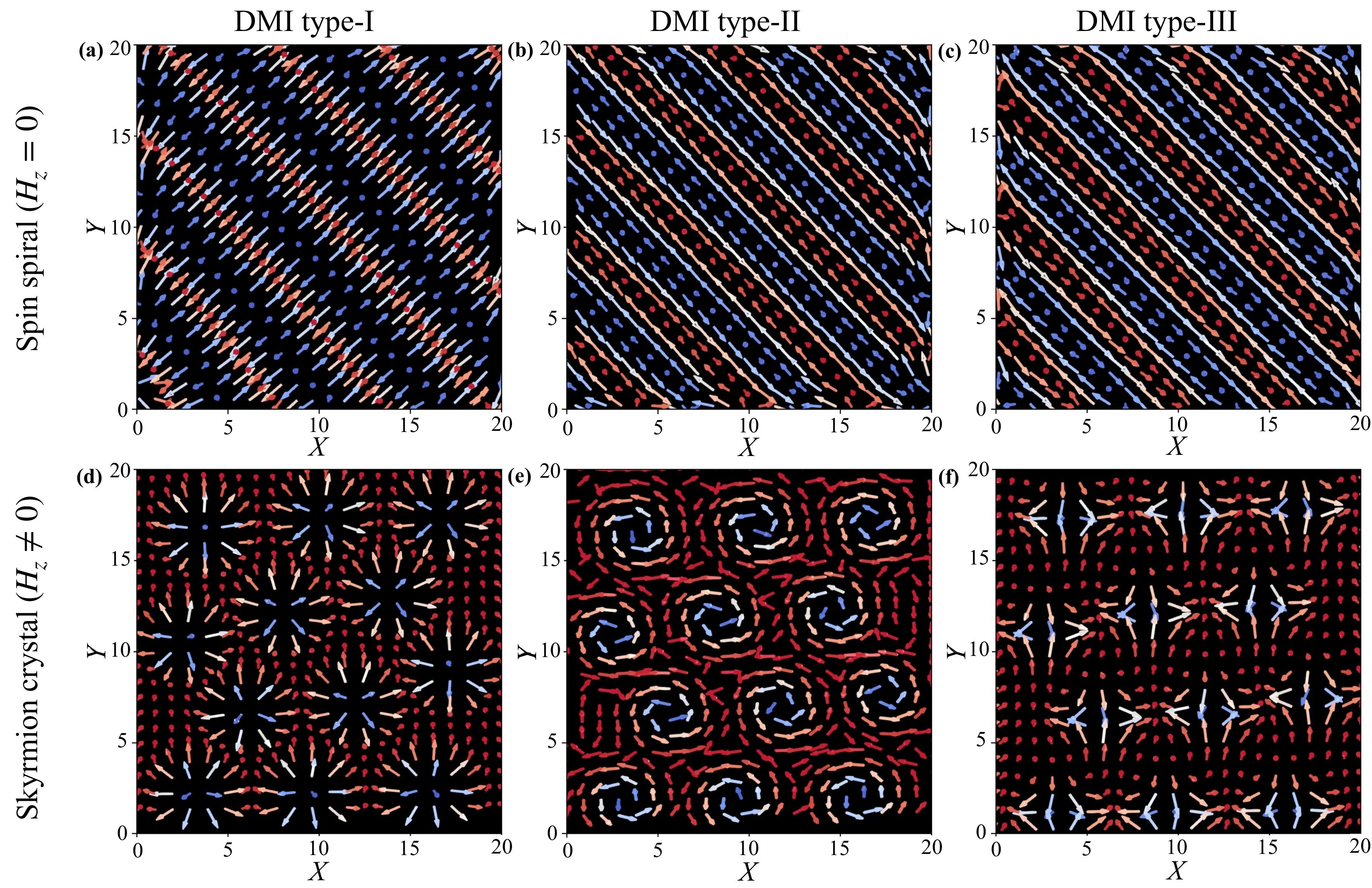,trim=0.0in 0.0in 0.0in 0.0in,clip=false, width=178mm}
\caption{Spin configurations in the spin spiral (SS) and the skyrmion crystal (SkX) phases, obtained in Monte Carlo annealing process at a low temperature $T\!=\!0.001J$ on a 24$\times$24 square lattice with periodic boundary conditions. (a), (b), (c) SS texture realized in the absence of any magnetic field ({\it i.e.} at $H_z\!=\!0$) for the three types of DMI discussed in Sec. II. (d),(e),(f) Triangular crystal structure of N\'eel, Bloch and anti skyrmions, respectively, obtained at a magnetic field $H_z\!=\!1.8J$. The colors in the arrows represent the $z$ component of the spin vector, blue (red) being the down (up) spin arrangement.  Other parameters used are $J\!=\!1$, $D\!=\!0.5J$, and $A\!=\!0.001J$.}
\label{fig2}
\vspace{-2mm}
\end{center}
\end{figure*}
\begin{figure*}[ht]
\begin{center}
\vspace{-0mm}
\epsfig{file=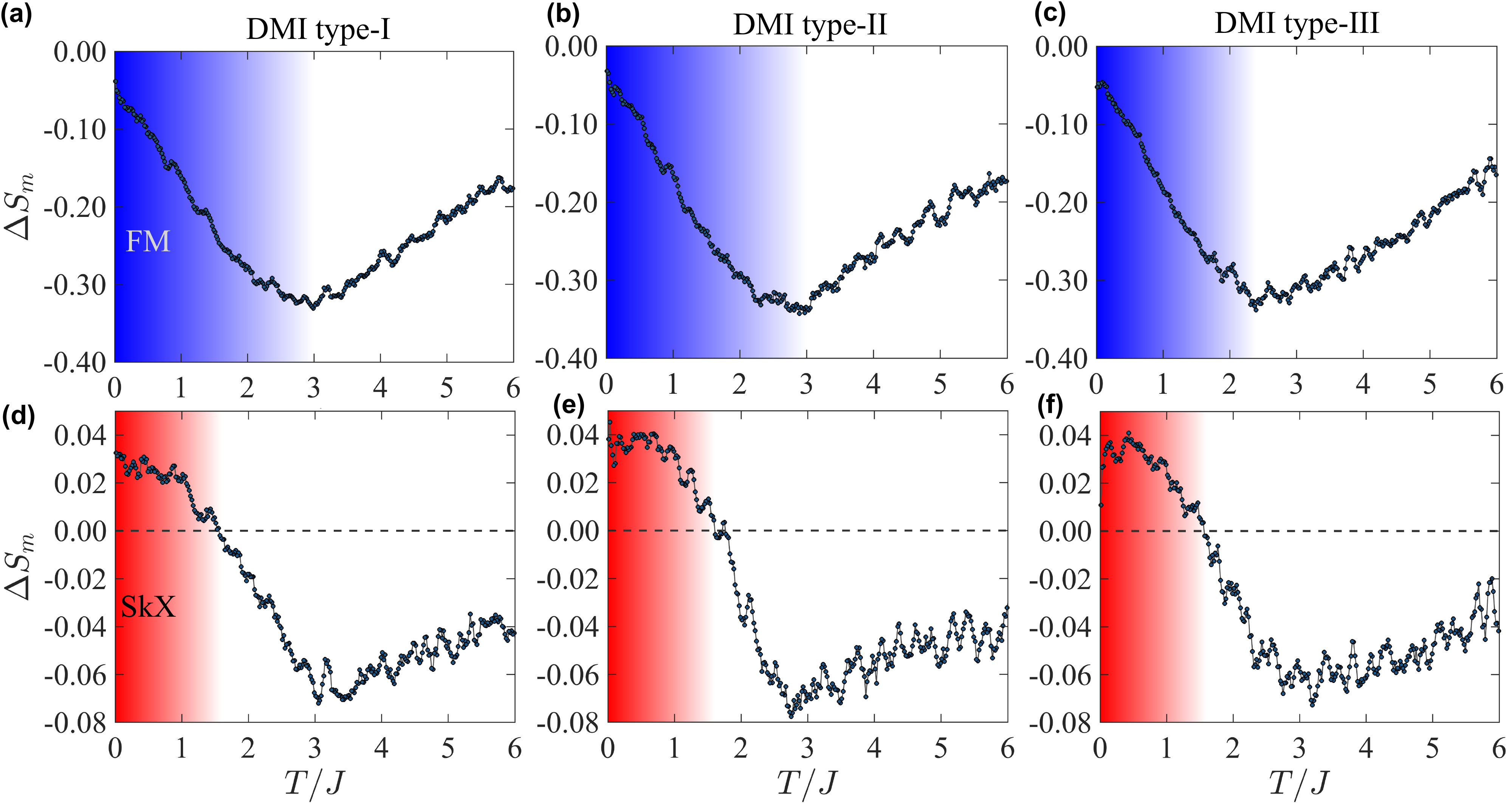,trim=0.0in 0.0in 0.0in 0.0in,clip=false, width=178mm}
\caption{(a),(b),(c) Temperature variation of magnetic entropy change $\Delta S_m$ for the three types of DMI discussed in Sec. II, at a magnetic field $H_z\!=\!4.4J$ which stabilizes the FM phase (blue region) at low temperatures. (d),(e),(f) Temperature variation of $\Delta S_m$ for the three types of DMI, at $H_z\!=\!1.8J$ which stabilizes the SkX phase (red region) at low temperatures. Other parameters used are $J\!=\!1$, $D\!=\!0.5J$, and $A\!=\!0.001J$. The upward turn in $\Delta S_m$ indicates the entrance to the ordered phase via the fluctuation-dominated regime. The change in the sign of $\Delta S_m$ occurs when the skyrmions stabilize and form a triangular crystal at low temperatures.}
\label{fig3}
\vspace{-2mm}
\end{center}
\end{figure*}
\begin{figure*}[ht]
\begin{center}
\vspace{-0mm}
\epsfig{file=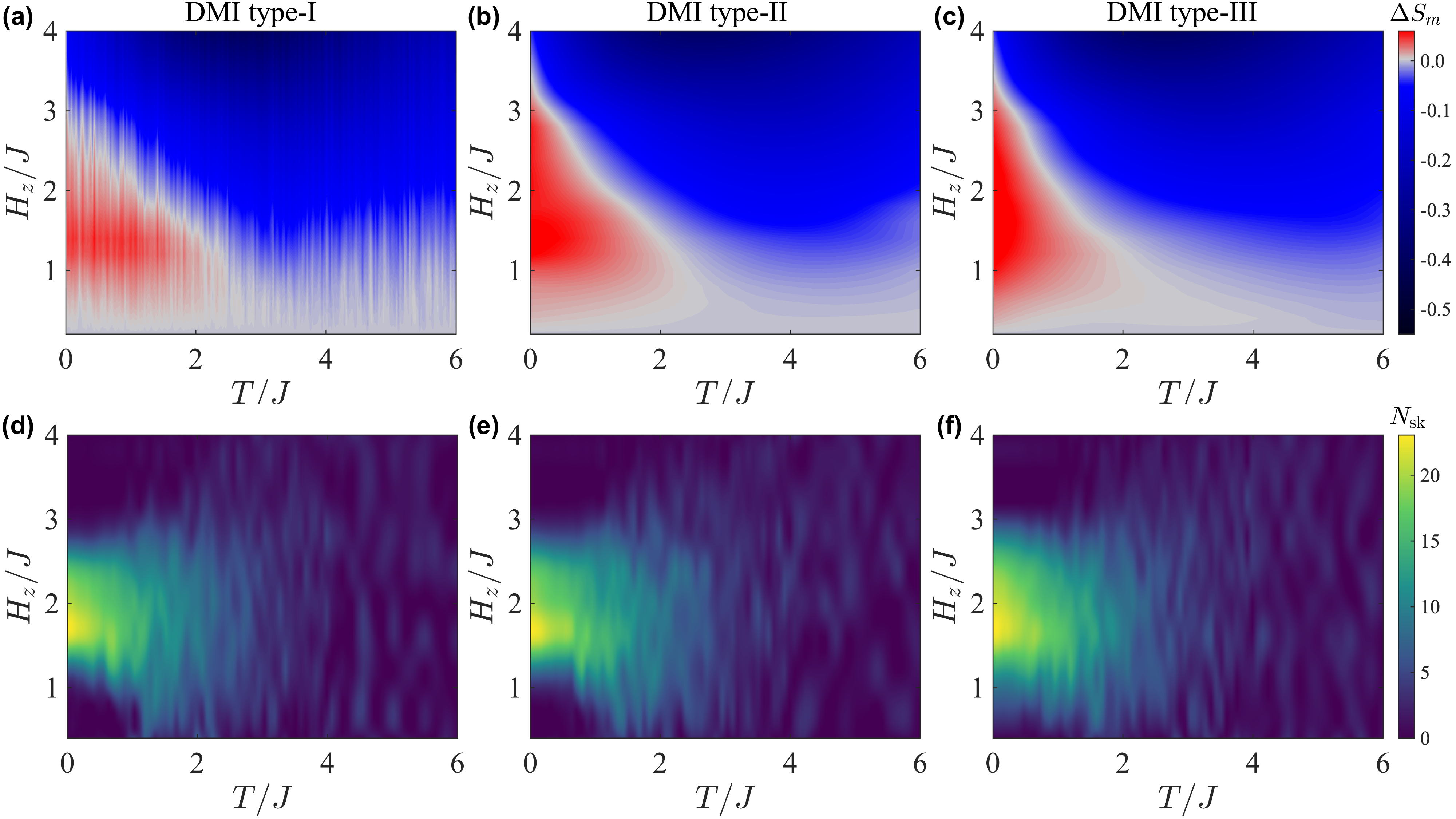,trim=0.0in 0.0in 0.0in 0.0in,clip=false, width=178mm}
\caption{Magnetic entropic change $\Delta S_m$ (top row in (a),(b),(c)) and the skyrmion number $N_{\rm sk}$ (bottom row in (d),(e),(f)) in the temperature $T$ vs. magnetic field $H_z$ plane, for the three types of DMI configurations. The top and bottom rows provide a magnetoentropic mapping of the skyrmion crystal phase---shown by the red regions in top panels and bright yellow regions in below panels. Parameters used are $J\!=\!1$, $D\!=\!0.5J$, and $A\!=\!0.001J$.}
\label{fig4}
\vspace{-2mm}
\end{center}
\end{figure*}


\noindent {\it Monte Carlo annealing:} Spin configurations are obtained using Monte Carlo (MC)
annealing procedure on a square lattice of size 24$\times$24 with open boundary conditions. The annealing process was started at a high temperature $T\!=\!10J$ with an initial random spin configuration, and the temperature was gradually lowered down to $T\!=\!0.001J$ in 500 steps. At each temperature step, $10^8$ MC spin update steps are performed. At each spin update step, a new direction of the spin vector at a randomly-selected site is chosen randomly within a small cone of angle $\Delta \theta=2^{\circ}$ around the initial spin direction. The new {\it trial} spin configuration is accepted or rejected using the standard metropolis algorithm: accepted if the energy difference $\Delta E$ between the two spin configurations is negative, and accepted with the Boltzmann probability $e^{-\Delta E/(k_BT)}$ when $\Delta E$ is positive; the latter is to incorporate the thermal spin fluctuation which increases with temperature.

\noindent {\it Calculation of magnetic entropy change:} The change in the magnetic entropy at a given temperature $T$ and magnetic field $H$ is obtained, using Eq.~(\ref{entropy_change}), as 
\begin{align}
\Delta S_m(H,T) = \frac{\mu_0}{T_2 - T_1} \int_0^H M(T_2,H)-M(T_1,H)  dH
\end{align}
where, $T_1\!=\!T-\Delta T$, $T_2=T+\Delta T$ are two considered temperatures around $T$, $\Delta T$ is incremental temperature gradient, $\mu_0$ is the free-space permeability, and $M$ is the average magnetization.
The above integral is evaluated numerically using the Simpson's method from the magnetization data obtained in MC simulations. To enhance the smoothness of the magnetic entropy variation with $T$, a Savitzky-Golay filter was applied. This filtering scheme uses a $4^{\rm th}$ order polynomial for local fitting, and reduces high-frequency noise from the entropy signal, revealing a clearer representation of the overall trend in the entropy change.

\noindent {\it Calculation of skyrmion number:} The skyrmion number in Eq.~(\ref{sk_num}) can be expressed in discretized form as
\begin{align}
N_{sk} = \frac{1}{4\pi}\sum_{i,j} {\bf S}_{ij}.[({\bf S}_{i+1,j}-{\bf S}_{i,j})\times({\bf S}_{i,j+1}-{\bf S}_{i,j})],
\end{align}
in which the integration in Eq.~(\ref{sk_num}) has been replaced by a summation and the partial derivatives are evaluated using a central-difference scheme. Magnetic skyrmions can be classified by the topological index $N_{sk}$, since under transformation of the underlying lattice from a torus to a sphere a skyrmion gives a full coverage of the sphere---a unique feature of the topological structure of the skyrmion. The number $N_{sk}$, therefore, helps in the characterization of the skyrmions, especially the SkX phase, in the phase plane spanned by the magnetic field  and temperature.



\section{Results} 
\subsection{Three types of skyrmion crystals}
In our two-dimensional spin systems, the SkX phase appears at low temperatures within a range of external magnetic fields, for all three types of DMI described above. The MC-generated spin configurations representing the chiral ordered phases, {\it viz.} the spin spiral (SS) which appear in the absence of any magnetic field and the SkX, for the three types of DMIs at a low temperature $T\!=\!0.001J$ are shown in Fig.~\ref{fig2}. Fig.~\ref{fig2}(a),(b),(c) show one of the two natural solutions of diagonal arrangements for the SS configurations. On the other hand, Fig.~\ref{fig2}(d),(e),(f) show triangular crystals of N\'eel, Bloch and anti skyrmions, respectively. Promising platforms in which such two-dimensional skyrmions can appear are interfaces between compounds with competing magnetic interactions and broken inversion symmetry at the interface~\cite{Balents_PRL2014,Yokouchi_JPSJ2015,Matsuno_SciAdv2016,Hoffmann2017,Nakamura_JPSJ2018,Vistoli_NPhys2019,Mohanta_PRB2019,Park_PRB2021}. At finite temperatures, the phase boundaries between the SS, the SkX and the fully-polarized FM phases are not sharp, but rather exhibit a region with metastable spin configurations with characteristics of the neighboring phases and governed primarily by thermal fluctuations~\cite{Mohanta_PRB2019}. One such metastable phase which is important for the below discussion is the precursor phase of the skyrmions which prevails at temperatures above the critical temperature for the SkX phase. 

\subsection{Temperature dependence of entropy change}
Fig.~\ref{fig3} shows the temperature dependence of the magnetic entropy change $\Delta S_m$ for the three types of DMI discussed in Sec. II and at two values of the magnetic field $H_z\!=\!4.4J$ and $H_z\!=\!1.8J$, at which the system is stabilized in the FM and the SkX phases at low temperatures, respectively, upon cooling from the spin-disordered paramagnetic phase. At $H_z\!=\!4.4J$ (Fig.~\ref{fig3}(a),(b),(c)), for which the FM phase is stabilized at low temperatures, the magnitude of $\Delta S_m$ is peaked at a critical temperature $T\!\approx\!3J$, at the onset of the FM phase, below which it starts to decrease. A similar behavior is observed at $H_z\!=\!1.8J$, for which the SkX phase is stabilized at low temperatures---$\Delta S_m$ shows upturn below a temperature $T\!\approx\!3J$. In addition, in this case $\Delta S_m$ changes sign at $T\!\approx\!1.8J$. This positive entropy change is revealed in all three types of the DMI settings.

\subsection{Entropy response in the skyrmion crystals}
To understand the connection between the features in the magnetic entropy change $\Delta S_m$ and various temperature dependent phase transitions, we compare $\Delta S_m$ with the skyrmion number $N_{sk}$ in the parameter plane spanned by temperature $T$ and the external magnetic field $H_z$, as shown in Fig.~\ref{fig4}. At larger values of the magnetic field ($H_z \!\gtrsim \!3J$), for which the fully-polarized FM phase is stabilized at low temperatures, $\Delta S_m$ varies continuously with temperature and exhibits a maximum in its absolute value near the critical temperature for the transition to the FM phase, as shown by the dark blue regions in Fig.~\ref{fig4}(a)-(c). At smaller values of the magnetic field ($H_z \!\lesssim \!0.5J$), the transition takes place from the spin-disoriented paramagnetic to the SS phase upon lowing the temperature. In this case, the change in the magnetic entropy is nearly zero, as shown by the white regions in Fig.~\ref{fig4}(a)-(c). At intermediate values of the magnetic field ($1.5\! \lesssim \! H_z \!\lesssim \!2.5J$), the SkX phase is stabilized at low temperatures, as evident from Fig.~\ref{fig4}(d)-(f). Remarkably, $\Delta S_m$ undergoes a sign change as the system enters into the SkX phase either from the paramagnetic phase or from the fully-polarized FM phase. The positive entropy change, shown by the red regions in Fig.~\ref{fig4}(a)-(c), is found for all three types of skyrmions realized in our two-dimensional systems, and also observed in three-dimensional materials in experiments ~\cite{Bocarsly_PRB2018,Dhital_PRB2020,Zuo_PRMater2021}.

Spin fluctuations play an important role in the temperature-driven phase transitions in Dyaloshinskii-Moriya magnets~\cite{Janoschek_PRB2013}. Magnetic entropy change, obtained via measurements of magnetization and specific heat, can provide a great amount of information about the nature of the fluctuation-dominated phase transitions~\cite{Bocarsly_PRB2018,Kautzsch_PRMat2020,Schueller_PRMat2020}. Both experimental and theoretical studies suggest that the temperature-driven transition from the paramagnetic to the SkX phase occurs via a precursor phase in which non-trivial topological spin textures of skyrmionic character are developed within a paramagnetic background with abundance of fluctuations~\cite{Wilhelm_PRL2011,Kindervater_PRX2019,MCSim}. This is evident from the plots of the skyrmion number $N_{\rm sk}$ in Fig.~\ref{fig4}(d)-(f), as this number fluctuates and its non-zero value is extended up to much higher temperature above the SkX phase, which prevails at temperatures $T\!\lesssim \!1.5J$. This fluctuation-dominated weak first order phase transition, also known as the Brazovskii transition, exists also at higher fields, as it can be seen from the magnetic entropy change in going from the paramagnetic to the ferromagnetic phase. The results presented in Fig.~\ref{fig3} and Fig.~\ref{fig4} enable us to conclude that the precursor phase involving skyrmions exists at temperatures $T\!\lesssim \!3J$, below which $\Delta S_m$ changes its slope. The essential features of $\Delta S_m$ are qualitatively the same for all three types of DMI considered in our models.

\section{Discussion and summary}
\begin{figure}[t!]
\begin{center}
\vspace{-0mm}
\epsfig{file=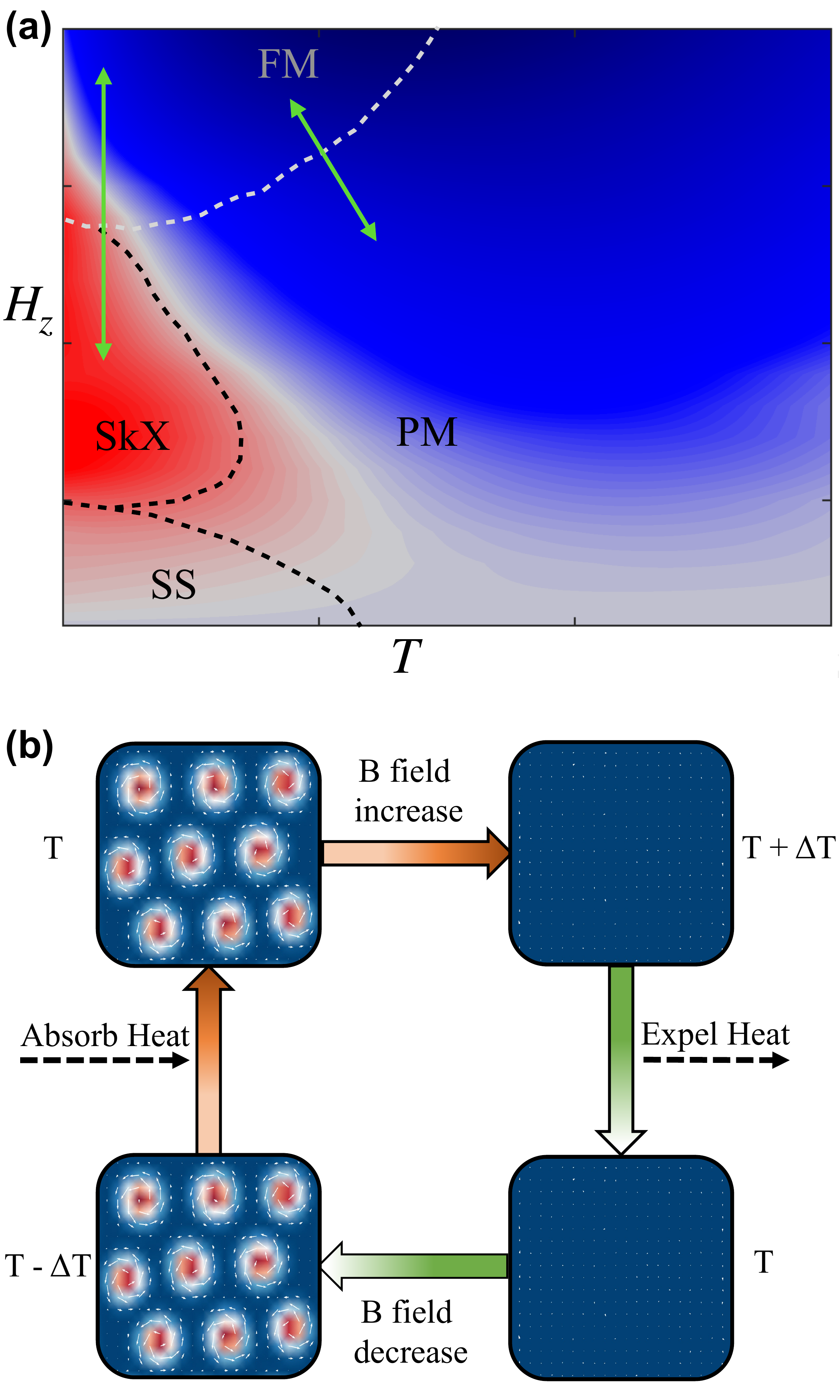,trim=0.0in 0.0in 0.0in 0.0in,clip=false, width=70mm}
\caption{(a) Schematic phase diagram, showing two routes of phase transition (green arrows) for magnetic cooling cycle: one between the paramagnetic (PM) and ferromagnetic (FM) phases, and another between the skyrmion crystal (SkX) and the FM phases. (b) Proposed cooling cycle between the SkX and the FM phases. Increasing magnetic field causes FM ordering, which results in increase in temperature of the material. Once excess heat is expelled out, the material is again demagnetized, and the temperature is decreased.}
\label{fig5}
\vspace{-5mm}
\end{center}
\end{figure}

Because of the sign reversal in the magnetic entropy change $\Delta S_m$ in the SkX phase, the fractional change in the magnetic entropy in transitioning from the fully-polarized FM phase to the SkX phase \textit{i.e.} $|\Delta S_m^{\rm SkX}-\Delta S_m^{\rm FM}|/\Delta S_m^{\rm FM}$ is always larger than the fractional change in the magnetic entropy in transitioning from the PM phase to the FM phase \textit{i.e.} $|\Delta S_m^{\rm FM}-\Delta S_m^{\rm PM}|/\Delta S_m^{\rm PM}$. In Fig.~\ref{fig5}(a), the two  routes for magnetic cooling process are shown; the PM to FM path is used in conventional magnetic cooling methods based on the magnetocaloric effect (MCE). The phase boundaries can be drawn explicitly by using various observables such as magnetization, spin-spin correlation function, syrmion number and topological Hall conductivity, as done in previous studies~\cite{Mohanta_PRB2019,Mohanta_PRB2020}.
The MCE also depends significantly on the nature of the variation in magnetization near the phase transitions. Based on the nature of the magnetic transition, a variety of magnetic materials have been proposed to enhance the efficiency of the MCE~\cite{Franco_annurev2012}. Materials with a first-order phase transition to the magnetically-ordered phase exhibit a large change in the magnetization over a small temperature range. It results in an abrupt change in the magnetic entropy, which leads to losses by hysteresis and irreversibilities of the magnetic state during demagnetization~\cite{Mellari_IJACR2023}. The energy lost during the cycle involving the magnetization and 
demagnetization, is dissipated as heat via the magnetic circuits. Since the Maxwell's relation in Eq.~(\ref{entropy_change}) is valid for systems in equilibrium, the dissipative kinetics essentially reduces the efficiency of the MCE of these materials. Additional structural transitions have also been reported in the vicinity of the first-order magnetic phase transition~\cite{Murtaza_PRB2020}. Materials with a smooth phase transition to the magnetically-ordered phase are, on the other hand, free from these penalizing phenomena such as hysteresis loss, slow kinetics and structural transition. Because of the above advantages, we propose materials hosting the SkX phase to be good candidates for the magnetocaloric applications. A typical cycle of the magnetic cooling method involving a SkX phase and a FM phase is shown in Fig.~\ref{fig5}(b). The proposed refrigeration cycle is performed by adiabatically magnetizing the system to FM phase from SkX phase, causing magnetic heating. This heat is then expelled through the system which again undergoes magnetic cooling through adiabtic demagnetization. Finally, the system absorbs heat from the environment and reaches its thermal equilibrium, from which it can restart the magnetic refrigeration cycle.


To summarize, we investigated the magnetic entropy change in two-dimensional magnets hosting various kinds of skyrmion crystals and the nature of fluctuation-dominated phase transitions. Our results confirm several experimental reports and suggest that the magnetic entropy change can be used as a general probe, as an useful alternative to neutron probes in momentum space~\cite{Muhlbauer_Science2009,Pappas_PRL2009,Kindervater_PRX2019}, transmission-electron microscopy imaging in real space~\cite{Yu_Nature2010,Yu_NMater2011,Tokunaga_NComm2015} or topological Hall effect-based probes~\cite{Yokouchi_JPSJ2015,Matsuno_SciAdv2016,Nakamura_JPSJ2018,Vistoli_NPhys2019}, to analyze the nature of phase transitions in Dzyaloshinskii-Moriya magnets in two and three dimensions. Our study also provides interesting perspectives about magnetocaloric effect in the SkX phase, and suggest that magnetic materials which host the SkX phase can be used in efficient magnetic cooling applications. More stable topological magnetic phases, such as hopfions which appear in three-dimensional materials, open up a new path towards the exploration of magnetocaloric applications with enhanced efficiency at elevated temperatures~\cite{Tokunaga_NatCommun2015,Kent2021,Rybakov2022}.

\section*{Acknowledgements} 
NM acknowledges support of an initiation grant (No. IlTR/SRIC/2116/FIG) from IIT Roorkee and SRG grant (No. SRG/2023/001188) from SERB. Numerical calculations were performed at the computing resources of PARAM Ganga at IIT Roorkee, provided by National Supercomputing Mission, implemented by C-DAC, and supported by the Ministry of Electronics and Information Technology and Department of Science and Technology, Government of India.


%

\end{document}